\documentclass[reprint,showpacs,preprintnumbers,amsmath,amssymb, aip]{revtex4-1}
\usepackage{graphicx,wasysym}
\usepackage{dcolumn}
\usepackage{bm}
\usepackage{natbib}
\usepackage{tabulary}
\usepackage{color}

\newcommand{\df}{d_\mathrm{f}}
\newcommand{\tauo}{\tau_{\mathrm{o}}}

\begin{document}

\preprint{1}

\title{The length and time scales of water's glass transitions}

\author{David T. Limmer}

 \email{dlimmer@princeton.edu}
\affiliation{%
Princeton Center for Theoretical Science, Princeton University, Princeton NJ 08540
}%

\date{\today}
\begin{abstract}
Using a general model for the equilibrium dynamics of supercooled liquids, I compute from molecular properties the emergent length and time scales that govern the nonequilibrium relaxation behavior of amorphous ice prepared by rapid cooling. Upon cooling, the liquid water falls out of equilibrium whereby the temperature dependence of its relaxation time is predicted to change from super-Arrhenius to Arrhenius. A consequence of this crossover is that the location of the apparent glass transition temperature depends logarithmically on cooling rate. Accompanying vitrification is the emergence of a dynamical length-scale, the size of which depends on the cooling rate and varies between angstroms and 10s of nanometers. While this protocol dependence clarifies a number of previous experimental observations for amorphous ice, the arguments are general and can be extended to other glass forming liquids.
\end{abstract}


\maketitle

Glasses of water, or amorphous ices, are nonequilibrium, nonergodic states of matter and as such their properties depend intimately on how they are prepared.\cite{binder2011glassy} As with other substances, glassy phases of water can be prepared by cooling the liquid, however due to the ease at which water crystallizes cooling rates required to vitrify water are much larger than those typically employed in other glassy-forming materials.\cite{bruggeller1980complete} Understanding how large cooling rates affect the properties of amorphous ice is of considerable interest because experimentally water must be cooled at rates greater than $10^8$ K/min, which is up to 8 orders of magnitude greater then those rates utilized when crystallization can be suppressed, such as in aqueous solutions and confined water.\cite{johari1987glass,capaccioli2011resolving,debenedetti2003supercooled,oguni2011calorimetric} Recently it has been shown that the reversible relaxation kinetics of supercooled water are well described by dynamic facilitation theory.\cite{C3FD00076A,limmer2013theory} By adopting recent extensions of this theory to systems arrested by cooling,\cite{keys2013calorimetric} the emergent length and time scales that result when liquid water is driven out-of-equilibrium can be computed. With these scales, the location of the water's glass transition temperature is shown to depend logarithmically on cooling rate, which is proposed to explain the broad range of previous experimental estimates.\cite{angell2004amorphous,velikov2001glass,oguni2011calorimetric,capaccioli2011resolving} 

Dynamic facilitation theory posits that below an onset temperature, $T_\mathrm{o}$, reorganization in a supercooled fluid is the result of localized excitations that exhibit no static correlations in equilibrium, but whose dynamics are highly correlated.\cite{chandler2010dynamics} At low temperatures where mobility is sparse, $T<T_\mathrm{o}$, the concentration of these excitations, $c$, is proportional to the Boltzmann factor,
\begin{equation}
\label{boltz}
c \propto \exp{\left [-(1/T-1/T_\mathrm{o} ) J_\sigma\right]} \, ,
\end{equation}
where $J_\sigma$ is the free energy for creating an excitation in the liquid with surrounding displacement $\sigma$. The motion of these excitations through the fluid is constrained, such that mobility is more likely to occur near regions that are already mobile and the direction of that mobility is preserved over a length scale greater than the size of the excitation. These kinetic constraints manifest themselves in dynamic heterogeneity, stretched exponential decay of time correlation functions  and transport decoupling.\cite{biroli2013perspective} The dynamics that result from these constraints are hierarchical,\cite{palmer1984models} with a logarithmic energy scaling relation for excitations accompanying different displacement lengths,
\begin{equation}
\label{gamma}
J_{a'}-J_a = \gamma J_\sigma \ln a'/a \, ,
\end{equation}
where $\gamma$ is a non-universal exponent of order unity that relates the likelihood of two excitations of different sizes meeting.   

The constituent relations in Eqs. \ref{boltz} and \ref{gamma} come from solvable lattice models such as the East\cite{jackle1991hierarchically} and Arrow models,\cite{garrahan2003coarse} but these basic behaviors also emerge in molecular dynamics simulations.\cite{keys2011excitations}  By coarse-graining short trajectories, in effect mapping molecular dynamics to these lattice models, the microscopic material properties $T_\mathrm{o}$, $J_\sigma$ and $\gamma$ can be determined. For water and water-like models at ambient pressure, these properties obey a principle of corresponding states such that the onset temperature is approximately given by the temperature of maximum density, experimentally corresponding to $T_\mathrm{o}\approx 277$~K, while in units of this temperature, $J_\sigma/T_\mathrm{o} = 23$ and $\gamma=0.63$.\cite{C3FD00076A} This microscopic procedure has so far only been applied to a few systems,\cite{keys2011excitations,C3FD00076A} however the macroscopic consequences of adopting this perspective have been more systematically studied. Among other things, the existence of localized excitations obeying these specified propagation rules determines the temperature dependence of the mean structural relaxation time. This form, known as the parabolic law, has been used to collapse experimental data on a wide variety of structural glass formers,\cite{elmatad2009corresponding,elmatad2010corresponding} including molecular dynamics models of water\cite{C3FD00076A} and nanoconfined water\cite{limmer2012phase} where complications to crystallization can be mitigated.

Below $T_\mathrm{o}$, the temperature dependence of the mean structural relaxation time is,
\begin{equation}
\label{tau}
\ln \tau(T)/ \tau_\mathrm{MF}= \left (1/T-1/T_\mathrm{o} \right ) J_\sigma \ln [\ell(T)/\sigma]^\gamma \, ,
\end{equation}
where at equilibrium the length $\ell(T)$ is given by, 
\begin{equation}
\label{leq}
\ell(T)/\sigma = e^{\left (1/T-1/T_\mathrm{o} \right ) J_\sigma/\df} \, .
\end{equation}
This length is simply the mean free path between uncorrelated excitations, whose concentration is determined by the Boltzmann factor in Eq.\,\ref{boltz}, and $\df$ is the fractal dimensionality of these paths, which has been determined numerically for three spatial dimensions to be $\df=2.6$.\cite{keys2011excitations} Substituting Eq.~\ref{leq} into Eq.~\ref{tau} gives the more familiar form of the parabolic law. Below $T_\mathrm{o}$, the super-Arrhenius temperature dependence observed for supercooled liquids can be seen as a consequence of the temperature dependence of this length that increases exponentially with decreasing temperature following the decreasing number density of excitations. The proportionality constant $\tau_\mathrm{MF}$ in Eq. \ref{tau} is the average time for attempting displacements above $T_\mathrm{o}$. While this time may have a temperature dependence itself, it is typically weak and for simplicity is assumed to be a constant, $\tau_\mathrm{MF} \approx 2.5 \times10^{-12}$ s.\cite{limmer2012phase} 

In order to extend this picture to irreversible relaxation, the protocol that drives the system out of equilibrium must be defined. In this work, only protocols that begin with the system in equilibrium and are cooled at constant rate with magnitude, $\nu = | \Delta T/\Delta t |$, to a temperature far below the glass transition are considered. The condition of beginning in equilibrium is critical, as it allows for the application of the theory for the reversible dynamics described above. The constant cooling rate and quench depth are chosen here for simplicity and can be relaxed to accommodate behaviors associated with annealing and aging. Accordingly, only the mean relaxation, or aging time, within the amorphous ice immediately following the cooling protocol is considered. Its temperature dependence  therefore implies the timescale for relaxation upon reheating the quenched material. 


Under a constant cooling rate, the condition for maintaining an equilibrium state is that the rate of change of the external temperature is slow relative to the corresponding change in the relaxation time of the system. This condition, or rather where it breaks down, defines the glass transition temperature, $T_\mathrm{g}$, where the system falls out-of-equilibrium. The identification of a single temperature is well defined only in-so-far as relaxation occurs in stages with well separated timescales. For the hierarchical dynamics considered here, this is exactly what happens. Specifically, as has been pointed out previously by Solich and Evans for the East model,\cite{sollich1999glassy} in the limit of vanishing concentration of excitations, relaxation of different domains occur in stages separated by $\mathcal{O}[\ell(T)]$. Therefore, on a timescale less than $\tau(T_\mathrm{g})$, only domains less than $\ell(T_\mathrm{g})$ are likely to relax, leaving excitations separated by distances greater than this length arrested for $T<T_\mathrm{g}$ and times less than $\tau(T_\mathrm{g})$. 

Given this time scale separation for the model dynamics considered here, $T_\mathrm{g}$ can be defined implicitly by the differential relationship,
\begin{equation}
\label{tx}
\frac{d \tau}{d T}\Big |_{T_\mathrm{g}} = \nu^{-1}\, .
\end{equation}
Assuming that $T_\mathrm{g}$ is on the order of $T_\mathrm{o}$, Eq. \ref{tx} can be evaluated and substituted into Eq.\,\ref{leq} to give the mean nonequilibrium distance between excitations, 
\begin{equation}
\label{lne}
\ell (T<T_\mathrm{g}) \approx \ell_\mathrm{ne} \approx \sigma \exp{\left[\sqrt{-\ln( \tilde{\nu})/\gamma \df }\right ]} \, ,
\end{equation}
where $\tilde{\nu}= 2 \nu \tauo J_\sigma^2 \gamma /d_\mathrm{f}T_\mathrm{o}^3$ is a dimensionless cooling rate. Following the arguments above, this length scale in the glassy phase is indicative of correlations generated in the material out-of-equilibrium that exist until the material is warmed. Unlike in equilibrium where the mean distance becomes increasingly large upon cooling, in the arrested state the length is frozen into the structure of the material and thus no longer changes appreciably with temperature. Correspondingly, below $T_\mathrm{g}$ the timescale of relaxation no longer has an additional temperature dependence from $\ell(T)$. Therefore in the amorphous ice, the relaxation changes from the parabolic form to an Arrhenius form.

\begin{figure}
\begin{center}
\includegraphics[width=8.cm]{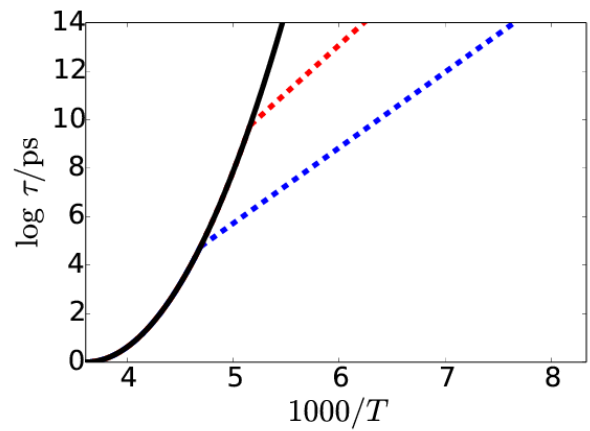}
\caption{Mean structural relaxation time, predicted for water as a function of temperature for different cooling rates:  1 K/min (black), 1$0^4$ K/min (red) and $10^7$ K/min (blue). Following Eq.~\ref{tx}. the location of the crossover to Arrhenius temperature dependence of $\tau$ indicates the glass transition temperature, $T_\mathrm{g}$. Following Eq.~\ref{tg}, locations where dashed lines increase to $\log \tau/\mathrm{ps}=14$ define the apparent glass transition temperature, $T_\mathrm{ag}$ as a function of cooling rate. While only the limiting behaviors at high and low temperature are directly observable for water, the implications of these mean timescales are pronounced, as discussed in the text.}
\label{Fi:1}
\end{center} 
\end{figure}

Figure \ref{Fi:1} shows the predicted relaxation time as a function of temperature for different cooling rates. As with all of the results presented in this paper, the parameters have been independently computed from microscopic calculations. The reversible liquid relaxation is predicted to be a smooth, monotonically increasing function given by Eq.~\ref{tau} and $\ell(T)$ in Eq.~\ref{leq}. The irreversible glassy relaxation is given by Eq.~\ref{tau} and $\ell_\mathrm{ne}$ in Eq.~\ref{lne} with $\tau_\mathrm{MF}$ determined by equating both curves at $T=T_\mathrm{g}$. These equations predict that for systems prepared by cooling rates that are slower than 1 K/min, follow the reversible curve over the timescales considered. Cooling rates that exceed 1 K/min however, exhibit a crossover to an Arrhenius temperature dependence. For faster rates, this crossover occurs at higher temperatures, corresponding to the system falling out of equilibrium from perturbations with shorter characteristic timescales. 

For bulk water, fast cooling rates are required due to the rapid rate of crystallization, therefore only indirect evidence of this crossover is observable. Nevertheless, a testable prediction of this theory follows from Eqs. \ref{tau}, \ref{leq} and \ref{lne} from which the change in slope at this crossover point can be evaluated. By construction, $\ell(T_\mathrm{g})=\ell_\mathrm{ne}$, so the change in slope of the relaxation time evaluated at $T_\mathrm{g}$, is invariant to cooling rate.  Though indirect for water, this crossover is observable for other glass forming liquids and for example is illustrated in Ref.~\onlinecite{wojnarowska2014deducting} complete with a change in slope of approximately $1/2$ as predicted from the above equations.  While the abruptness of the crossover is a product of assuming the relaxation time is given by a single length scale, corrections to this approximation will certainly smooth the crossover. However, to the extent that the distribution of lengths after the cooling protocol is sharply peaked around its mean value, as has been shown in Ref.~\onlinecite{keys2013calorimetric}, then for $T<T_\mathrm{g}$,  $\tau(T)$ will be well determined by this mean field approximation. For water, while the crossover itself is not observable, both the higher temperature behavior of the liquid and the low temperature behavior of the amorphous ice can be observed and be used to verify this prediction.

The changing temperature dependence in $\tau(T)$ is a consequence of falling out of equilibrium, arresting the liquid and forming a glass. As such it is expected to be a general feature of supercooled liquids driven from equilibrium and thus distinct from observations of crossovers noted previously for water under nano-confinement.\cite{bertrand2013deeply} Observations in confined water are made from nominally equilibrium measurements in pores whose disordered surfaces and sizes frustrate crystallization.\cite{limmer2012phase} The specific origin of this behavior is the subject of debate,\cite{bertrand2013deeply} however previous work suggests that it is due to freezing in confinement.\cite{limmer2012phase}  The behavior detailed here is, however, related to previously inferred transitions that have been invoked to either explain the weak thermal response upon heating hyper-quenched amorphous ice\cite{angell2008insights} or the variability in temperature dependence of the relaxation time at the glass transition.\cite{kivelson2001h2o} Indeed, crossovers of this sort have been previously postulated to reconcile the location of water's $T_\mathrm{g}$ with fits to the equilibrium liquid relaxation time,\cite{smith1999existence} however in this work the crossoever is explicitly considered a nonequilubrium effect, not predicted to occur in the reversible behavior of liquid water.

\begin{figure}
\begin{center}
\includegraphics[width=8.cm]{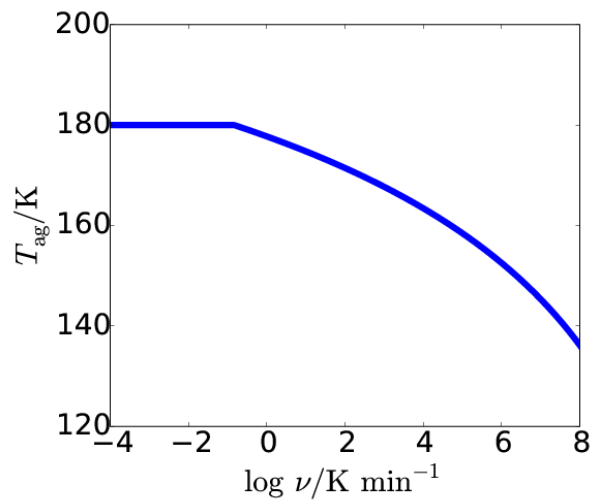}
\caption{Apparent glass transition temperature of water upon heating, as a function of cooling rate, defined in Eq.~\ref{tg}. The crossover near cooling rates of 1K/min occurs as the $T_\mathrm{ag}$ becomes equal to $T_\mathrm{g}$ as defined in Eq.~\ref{tx}. Note that experimentally only $\nu > 10^7 \mathrm{K/min}$  result in direct vitrification of water.}
\label{Fi:2}
\end{center} 
\end{figure}

Because of the crossover in temperature dependence of the mean relaxation time, there exists a temperature below  $T_\mathrm{g}$ where the system will begin to relax on laboratory timescales. This transition to a softening glassy state, is denoted  
the apparent glass transition temperature, $T_\mathrm{ag}$. In water this transition is often followed by abrupt crystallization, which does not allow for the complete recovery of the liquid. The apparent glass transition temperature is defined by, $\tau(T_\mathrm{ag}) = 100\, \mathrm{s}$, or for $\nu>1\,\mathrm{K}/\mathrm{min}$,
\begin{equation}
\label{tg}
 T_\mathrm{ag}^{-1} = T_\mathrm{o}^{-1} + \ln(100\mathrm{s}/\tau_\mathrm{MF})/  J_\sigma \sqrt{-\ln( \tilde{\nu}) \gamma/ \df } \, ,
 \end{equation}
otherwise $T_\mathrm{ag}$ equals the temperature where the extrapolated liquid relaxation time reaches 100s, which for water is about 180 K.\cite{C3FD00076A} While, Eq.~\ref{tg} refers explicitly to where relaxation takes place on timescales of 100s this can be generalized.\cite{limmer2013theory} Figure \ref{Fi:2} plots the apparent glass transition temperature which is shown to depend logarithmically on cooling rate. This dependence is irrelevant to many experiments on glass forming liquids, as cooling rates usually range between 1-10 K/min. However, for water, this dependence is crucial because cooling rates greater than $10^8$K/min are used in order to bypass crystallization. Cooling rates that are less than 1 K/min do not change $T_\mathrm{ag}$, as these perturbations would cause the system to fall out of equilibrium at larger characteristic timescales than 100s. However, cooling rates greater than 1 K/min are expected to affect the observed $T_\mathrm{ag}$ systematically, with changes of over 50 K expected for cooling rates that vary between $10 - 10^8$K/min.  Similar logarithmic cooling rate dependences has been noted experimentally for other network-forming liquids,\cite{moynihan1974dependence,bruning1992glass} and in simulations,\cite{vollmayr1996cooling} however this work seems to be the first to derive such a dependence from microscopic properties. This distinction between $T_\mathrm{g}$ and $T_\mathrm{ag}$ has been noted previously by Angell, who called this softening and its associated thermal response a ``shadow glass transition," however this work is the first to give a quantifiable mechanistic explanation.\cite{yue2004clarifying}


The logarithmic dependence of $T_\mathrm{ag}$ for water on cooling rate explains the apparent contradiction between dielectric measurements and extrapolated calorimetry measurements.\cite{capaccioli2011resolving} Dielectric measurements directly probe the timescale for molecular reorganization and designate the location of the glass transition temperature according to the definition in Eq.~\ref{tg}. Accordingly, estimates for the location of the glass transition with samples cooled at rates between $10^7 - 10^8$K/min yield values near 130-140K in agreement with the results for $T_\mathrm{ag}$ presented here.\cite{johari1987glass} Alternatively, calorimetry measurements probe the recovery of the liquid upon heating from the glass and designate the temperature where the system falls out of equilibrium as the location of the glass transition, corresponding to the definition of $T_\mathrm{g}$ in Eq.~\ref{tx}. While these measurements are not directly possible in bulk water as crystallization interferes, extrapolations using data for aqueous solutions\cite{velikov2001glass} and measurements of water in confinement\cite{oguni2011calorimetric} both yield values for $T_\mathrm{g}$ between 160-180K. While extrapolation from bulk water behavior to water in confinement or concentrated solution is not straightforward, previous work has shown that in many instances effects from these perturbations result in small corrections to $J_\sigma,\, T_\mathrm{o}$ and $\tau_\mathrm{MF}$.\cite{limmer2012phase} Figures \ref{Fi:1} and \ref{Fi:2}, show that it is only when cooling rates are on the order of 1 K/min 
are expected to yield $T_\mathrm{g}\approx T_\mathrm{ag}$. 

This argument against a value of $T_\mathrm{g}$ for water near 140 K presumes that observations of a weak thermal response for hyper-quenched or otherwise prepared amorphous ice near that temperature is not indicative of a recovering liquid state, but rather is a nonequilibrium response to impending crystallization. The smallness of the observed heat capacity relative to the known equilibrium liquid value\cite{kohl2005water} suggests that this is a reasonable perspective, nevertheless it stands as an implicit assumption to the interpretation of this theory as applied to water.  Note however that the definition of  $T_\mathrm{g}$ employed here is where the liquid falls out of equilibrium, which in order to bypass crystallization must occur at temperatures greater the the limit of homogeneous nucleation, 220 K.\cite{debenedetti2003supercooled}
 
\begin{figure}
\begin{center}
\includegraphics[width=8.cm]{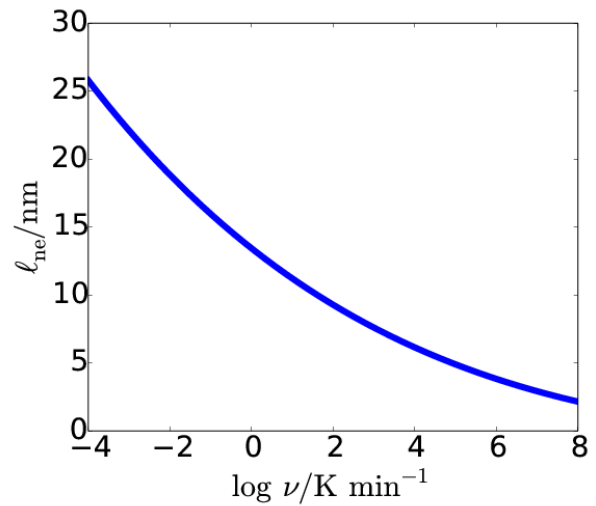}
\caption{Mean distance between excitations locked into the structure of amorphous ice prepared at different cooling rates. The nonequiliubrium correlation length determines the free energy barrier to relaxation in the glass. Note experimentally only $\nu>10^7\,\mathrm{K/min}$ result in direct vitrification of water.}
\label{Fi:3}
\end{center} 
\end{figure}

The nonequilibrium length defined in Eq.~\ref{lne} is plotted in Fig.~\ref{Fi:3} with $\sigma=2.5\mathrm{\AA}$. While generated by a intrinsically dynamic process, this length is a static measure of correlations in the arrested phase and as such similar to other lengths previously discovered in arrested amorphous matter.\cite{marcotte2012nonequilibrium}  Apart from determining the mean relaxation time, the magnitude of this length gives a geometrical explanation on the stability of a glass.\cite{keys2013calorimetric} For a given temperature, glasses produced with slower cooling rates relax on exponentially longer times as their excitations must overcome ever-increasing distances in order to relax. The weak thermal response upon heating through $T_\mathrm{ag}$ that hyperquenched water exhibits \cite{kohl2005water} is a manifestation of the weak correlations between excitations that result under such rapid cooling protocols, which when overcome in hyper-quenched water result in crystallization. Such a dependence of the thermal response on this correlation length and the corresponding scaling of this correlation length with cooling rate should be observable by small angle x-ray scattering experiments in amorphous ice at low temperatures, prepared with different cooling rates. These correlations, which likely result in additional scattering intensity at wave vectors inversely proportional to $\ell_\mathrm{ne}$, are left as a verifiable prediction of the microscopic theory applied here. Cooling rates in excess of $10^6\,\mathrm{K/min}$ must be used to directly vitrify water, thus other protocols must be used in order to generator amorphous ices with larger nonequiliubrium lengths. While a limited range of cooling rates for water are practically available experimentally, in principle this dependence of length scale on cooling rate is generic to glass forming systems and could be realized directly in aqueous solutions, albeit with different $J_\sigma,\,T_\mathrm{o}. \gamma$ and $\tau_\mathrm{MF}$.

In this work, the distribution of excitations is treated in a mean field way with one dominate length scale determining relaxation. More complicated nonequilibrium protocols such as pressure amorphization, annealing and polyamorphic transitions, likely generate distributions of excitations characterized by significantly different correlations. By preparing a material with a more complex protocol it is therefore possible that multiple length scales might be required for an adequate description of relaxation. 
For instance, the significant pressure dependence of $J_\sigma$\cite{limmer2013theory} dictates that amorphous ice produced at high pressure is characterized by a nonequilibrium length that is different from a corresponding sample of amorphous ice produced at low pressure. Depending on the relative ordering of these lengths, and if they are significantly separated, relaxation of a high pressure amorphous ice heated at low pressure could require a description of relaxation that incorporates both length scales as typical thermal fluctuations could relax regions characterized by the shorter of these two length at lower temperatures than required to relax the entire system. Such a construction can potentially explain observations by Amann-Winkel, et al~\cite{amann2013water} of multi-step relaxation behavior. \footnote{While this work was under review, Ref.~\onlinecite{limmer2013theory} demonstrated that this suggestion yields accurate results, as determined by direct simulation and compared with experiment.} 

This work presents expressions describing the nonlinear response for the simplest nonequilibrium protocol applied to a supercooled liquid: cooling with a specified constant rate. As alluded to above, many of the simplifications made for algebraic convenience are not necessary for numerical computations. Indeed, the consequences of annealing and aging can be taken into account with simple numerical calculations.\cite{keys2013aging}  By combining these expressions with the theory for thermal response provided in Ref.~\onlinecite{keys2013calorimetric}, the consequences of hyper-quenching rate on the response observed by calorimetry can be analyzed. That specific study is left for future work. 

Though the focus of this work is on the relaxation behavior of water, where the parameters entering that theory have been determined independently  from microscopic principles without adjustment, the general phenomena accompanying driving a liquid into an amorphous arrested state are generic. Supercooled liquids relax with an ever decreasing rate as the temperature is lowered, with dynamics that are hierarchical. When their structure is arrested any configurationally rare region that allows for relaxation becomes frozen in.  Therefore, it must be that irreversible relaxation in such an otherwise non-reorganizing material be governed by a temperature independent energy scale.

\acknowledgments
We would like to thank Aaron Keys and David Chandler for helpful discussions. This research was supported by the Princeton Center for Theoretical Science.

%


\end{document}